\documentclass[journal=jcpl,email=jornada@stanford.edu]{achemso}

\usepackage[table,xcdraw]{xcolor}

\usepackage{amsmath}
\usepackage{amsthm}
\usepackage{xcolor}

\usepackage{tabularx} 
\newcolumntype{Y}{>{\centering\arraybackslash}X}
\newcolumntype{B}{>{\centering\columnwidth=2\columnwidth}Y}
\newcolumntype{S}{>{\centering\columnwidth=.75\columnwidth}Y}

\usepackage{graphicx}
\usepackage{times}

\usepackage{soul}
\usepackage{color} 
\usepackage{physics}
\usepackage{mathtools}
\usepackage{siunitx}

\usepackage{braket}

\newlength{\singlecolumn}
\newcommand{\VSB}{\ensuremath{V_{S\rightarrow B}}}

\definecolor{mygreen}{RGB}{70, 170, 70}

\makeatletter
\renewcommand*{\@fnsymbol}[1]{\ensuremath{\ifcase#1\or \dagger\or *\or  \ddagger\or
\mathsection\or \mathparagraph\or \|\or **\or \dagger\dagger
\or \ddagger\ddagger \else\@ctrerr\fi}}
\makeatother

\usepackage{hyperref}
\hypersetup{
    colorlinks,
    citecolor=black,
    filecolor=black,
    linkcolor=black,
    urlcolor=black
}

\usepackage{graphicx}
\usepackage{changepage}

\usepackage{dirtytalk}
\usepackage{amstext} 
\usepackage{array}   

\usepackage{amssymb}
\usepackage{textcomp}
\usepackage{bm}

\usepackage{mathtools}
\newcommand\ddfrac[2]{\frac{\displaystyle #1}{\displaystyle #2}}

\newcommand{\heff}{h^{\mathrm{eff}}}

\newcommand{\dV}{\ensuremath{\Delta V}}

\newcommand{\vectau}{\ensuremath{\bm{\tau}}}
\newcommand{\dtau}{\ensuremath{\Delta\bm{\tau}}}
\theoremstyle{definition}

\title{Identifying Hidden Intracell Symmetries in Molecular Crystals and their Impact for Multiexciton Generation}

\author{Aaron R. Altman}
\affiliation{Department of Materials Science and Engineering, Stanford University, Stanford, CA 94305, USA}

\author{Sivan Refaely-Abramson}

\affiliation{Department of Molecular Chemistry and Materials Science, Weizmann Institute of Science, Rehovot 7610001, Israel}

\author{Felipe H. da Jornada}
\email{jornada@stanford.edu}
\affiliation{Department of Materials Science and Engineering, Stanford University, Stanford, CA 94305, USA}

\begin{document}

\begin{abstract} 
Organic molecular crystals are appealing for next-generation optoelectronic applications, most notably due to their multiexciton generation process that can increase the efficiency of photovoltaic devices. However, a general understanding of how crystal structures affects multiexciton generation processes is lacking, requiring computationally demanding calculations for each material.
Here we present an approach to understand and classify such crystals and elucidate multiexciton processes. We show that organic crystals that are composed of two sublattices are well-approximated by effective fictitious systems of higher translational symmetry.
Within this framework, we derive hidden selection rules in crystal pentacene and predict that the common bulk polymorph supports fast Coulomb-mediated singlet fission about one order of magnitude more than the thin-film polymorph, a result confirmed with many-body perturbation theory calculations.
Our approach is fully based on density-functional theory calculations, and provides design principles for the experimental and computational discovery of new materials with efficient non-radiative exciton decay rates.
\end{abstract}

\maketitle

Acene-based organic molecular crystals\cite{Kitaigorodsky1973} possess unique properties for photogeneration applications\cite{Kronik2016, Sharifzadeh2018, Forrest2007}, such as relatively large quasiparticle gaps, strongly bound electron-hole pairs (with exciton binding energies as large as few hundreds of meV), and sizable exchange interactions, leading to significant energy differences between spin singlet and triplet exciton states\cite{Cudazzo2012, Sharifzadeh2013, Rangel2016}. The latter allows for the unique phenomena of singlet fission, where optically excited spin-singlet excitons rapidly decay into pairs of spin-triplet excitons\cite{Smith2013, Rao2017, korovina2020lessons, Berkelbach2017}. A well studied example is solid pentacene, where experiments can separately identify the singlet-to-bitriplet decay rate from time-resolved spectroscopy and two-photon photoemission\cite{Rao2010, Wilson2011, Yost2014, Yong2017, Rao2017, tempelaar2018vibronic, duan2020intermolecular}, and the spatial dissociation and spin decoherence of the bitriplet state\cite{Bakulin2016, Yong2017}. As singlet fission is a promising path to increase the energy conversion yield in photovoltaic devices \cite{gish2019emerging, zhu2016two}, there is great interest to study the underlying exciton-biexciton decay mechanism and the effect of crystal structure on the exciton-exciton interactions\cite{qiu2021signatures}. While recent studies reveal direct relations between crystal packing and fission\cite{Margulies2017,Margulies2017b, Le2018, Buchanan2019, zaykov2019singlet, Coto2014, johnson2013role, hestand2015exciton}, computational predictions are critical to set design rules in a systematic way.

Theoretical approaches of singlet fission typically employ correlated wavefunction-based methods, such as configuration interaction (CI)\cite{Yost2014, Berkelbach2014, Berkelbach2017, Smith2013, Rao2017, Monahan2015, Casanova2018}, modelling the phenomenon using molecular dimers or small clusters. In fact, dimers are often considered good proxy models for singlet fission, though different studies debate whether a slip-stacked or parallel dimer orientation was preferable for efficient singlet fission coupling\cite{Havlas2016, Buchanan2017, Buchanan2017b, Smith2010, Feng2013}. However, these finite models do not naturally capture the periodic crystal environment, and hence are not the natural framework to elucidate exciton coupling as a function of crystal packing and symmetry

\cite{Buchanan2019, Le2018}. Indeed, recent studies suggest that the crystal symmetry stemming from its solid-state packing plays a key role in non-radiative exciton decay mechanisms\cite{Sharifzadeh2013, Refaely-Abramson2017, Monahan2015, Berkelbach2014,Sharifzadeh2018}, and interacting Green's-function based approaches offer a complementary method that naturally captures the periodic crystal environment\cite{Refaely-Abramson2017}. Nevertheless, such methods are not as mature as quantum-chemistry-based approaches. In particular, it is highly desirable to obtain simple design rules for such multiexciton processes in molecular crystals from straightforward density-functional theory (DFT) calculations and connect them to the previously mentioned dimer picture.

This work addresses these questions and reveals critical selection rules for exciton dynamics that are associated with the effective symmetry of the frontier bands of the systems. Our approach constitutes two main advances: first, we show how to construct an \emph{effective, high-symmetry tight-binding Hamiltonian} that accurately reproduces the low-energy band structure of the real material. This general effective Hamiltonian does not involve the real orbitals of the system, such as those obtained from maximally localized Wannier functions (MLWFs), and easily explains important features in the DFT band structure. Second, we show how to derive simple selection rules on excitonic processes based on these inexpensive DFT calculations and effective symmetry analysis. We demonstrate the method by considering two polymorphs of pentacence -- bulk and thin film -- and computing the singlet--bitriplet coupling matrix elements using explicit many-body perturbation theory calculations and a new approach to symmetry analysis. Our crystal-symmetry analysis correctly reproduces the fully \textit{ab initio} results, which predict that the thin film polymorph should not display fast singlet fission mediated by Coulomb interactions. Moreover, the methods used here are not specific to singlet fission, and can be extended to other excitonic processes satisfying some base assumptions. Our findings provide an understanding of singlet fission as a result of the details of the crystal structure, and computationally inexpensive tests for ongoing high-throughput computational efforts searching for novel materials for energy conversion~\cite{perkinson2019discovery, blaskovits2020designing, wang2019phenylated, einzinger2019sensitization, padula2019singlet, wang2018possibility}.

\begin{figure}[!htb]
    \centering
    \includegraphics[width= \singlecolumn]{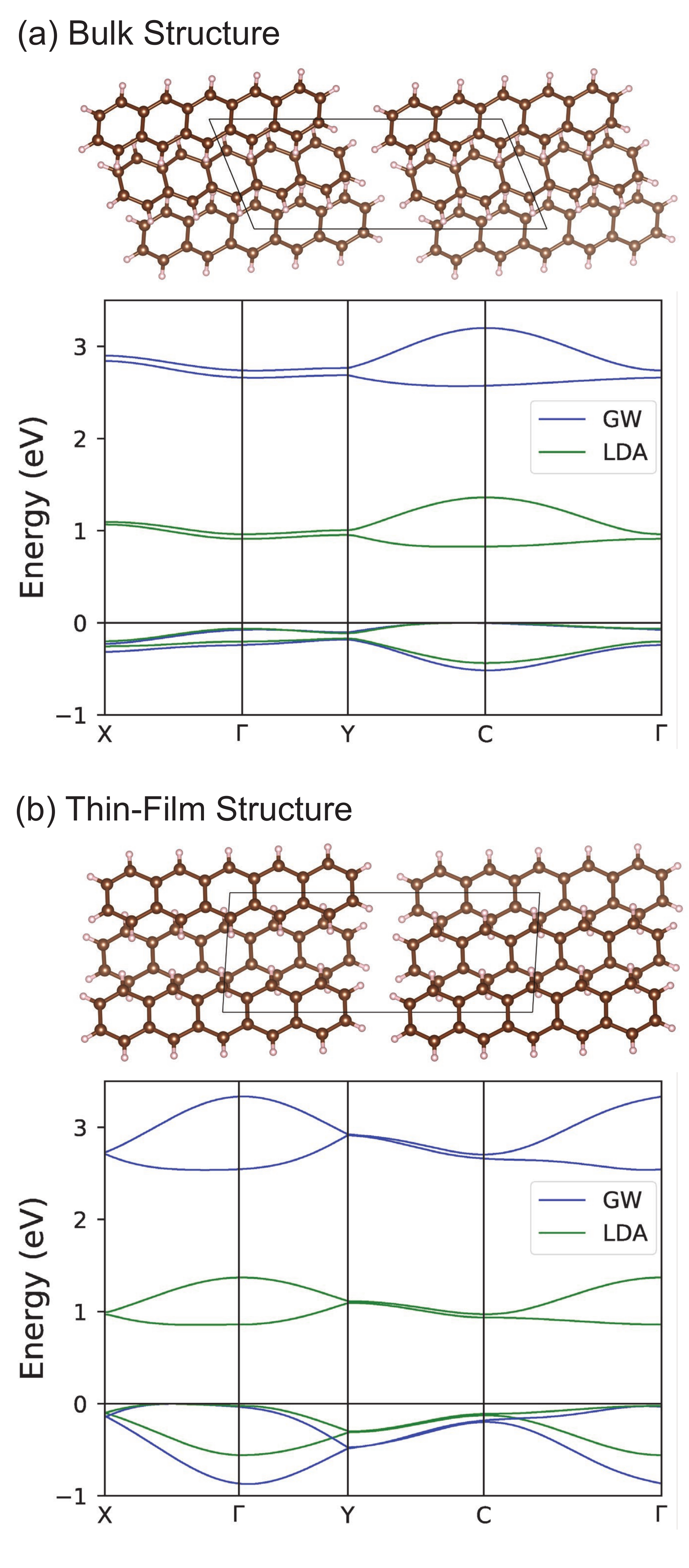}
    \caption{\label{fig:structure_bs} Crystal and band structure of (a) bulk and (b)
    thin-film pentacene, where the unit cell of both crystal structures is outlined by the black box. Band structures computed at both DFT and GW levels, and show the first two valence and conduction bands. The main features of the GW-corrected quasiparticle band structure are similar to the DFT ones, except for the value of band gap and Davydov splitting.}
\end{figure}

We begin by elucidating the similarities between the polymorphs of pentacene we study in this work, bulk and thin-film. Both polymorphs are structurally similar: they contain two nonequivalent molecules per unit cell and display crystal inversion symmetry. As shown in Fig.~\ref{fig:structure_bs}, the main difference in their crystal structures is the slip-stacking in bulk. Their electronic band structure is also very similar, except that the maximum band splitting in the frontier valence or conduction bands in one polymorph occurs approximately at the wavevector for which there is minimum splitting in the other. Aside from this difference, their electronic properties are similar -- including band gaps, Davydov splittings, and effective masses -- both when treated at a mean-field DFT level and when quasiparticle corrections are accounted for within the GW approximation~\cite{Hybertsen1986}. Additionally, both polymorphs are quasi-2D, which we verify by fictitiously increasing the out-of-plane lattice constant and observe minimal changes in the DFT band structure.

The approach we propose to analyze such classes of polymorphs is to define an effective, higher-symmetry Hamiltonian $\heff$ that reproduces the low-energy quasiparticle band structure (DFT or GW) and single-particle optical matrix elements at the expense of not reproducing the true wavefunctions of the underlying molecules. Importantly, for the prototypical molecular crystals we are interested in, $\heff$ is defined on a lattice that has only \emph{one effective site} per unit cell, even though it approximately describes the same low-energy physics as the real crystal containing two molecules per unit cell. 

We first compute the tight-binding Hamiltonian $H$ for the molecular crystal defined in its proper structure and unit cell: we simply compute the DFT band structure for the relevant polymorph and rotate it to a localized basis of isolated frontier molecular orbitals~\cite{pizzi2020wannier90}. To describe singlet fission and reproduce the DFT band structure, we require only the highest occupied molecular orbital (HOMO) and lowest unoccupied molecular orbital (LUMO) of each molecule (see SI). We directly use the frontier orbitals for our tight-binding basis of $H$ instead of employing MLWFs, since the latter display a slightly lower symmetry. However, our procedure is not sensitive to the choice of localized orbitals.

We denote the tight-binding matrix elements of $H$ between an orbital $\chi_i$ localized at the molecule site $\vectau$ and an orbital $\chi_j$ at $\vectau' + \vb{R}$ by
\begin{equation}
    \label{eqn:h_ij}
    h_{ij}^{\vectau}(\dtau) \equiv \braket{\chi_i(\vb{r}-\vectau)|H|\chi_j(\vb{r}-\vectau'-\vb{R})},
\end{equation}
where $\vb{R}$ is a lattice vector, $\vectau$ and $\vectau'$ are basis vectors, and $\dtau \equiv \vectau' + \vb{R} - \vectau$ is a vector between any two localized molecular sites. We emphasize that $\dtau$ is not a lattice vector. For both polymorphs of interest, there are two possible values of $\vectau$ in high-symmetry positions that label two distinct molecular sites in the unit cell: $\vectau_1=\vb{0}$ and $\vectau_2=\frac{1}{2}\vb{a}_1 + \frac{1}{2}\vb{a}_2$, where $\vb{a}_i$ are the primitive lattice vectors.

Consistent with the structural sublattice symmetry, we find that the hopping matrix elements between distinct sites are approximately the same for a fixed translation vector $\dtau$ between the initial and final orbital, irrespective of the initial site $\vectau$. This is demonstrated for the LUMO $\rightarrow$ LUMO hoppings in Table \ref{table:LUMO_hops} (see SI for details), showing near-identical values of $t_i\equiv h_\mathrm{LUMO,LUMO}^{\vectau_1}(\Delta\vb{r}_i)$ and $t_i'\equiv h_\mathrm{LUMO,LUMO}^{\vectau_2}(\Delta\vb{r}_i)$, where $\Delta\vb{r}_1=\frac{1}{2}\vb{a}_1 + \frac{1}{2}\vb{a}_2$ and $\Delta\vb{r}_2=\frac{1}{2}\vb{a}_1 - \frac{1}{2}\vb{a}_2$ (see Fig.~\ref{fig:fictitious} (a) and (d)).

\bgroup
\def\arraystretch{1.5}%
\begin{table}[h]
\caption{LUMO-to-LUMO hopping matrix elements, in meV, following the label convention in Fig.~\ref{fig:fictitious}. Note that $t$ and $t'$ only differ by the sublattice index (see Fig.~\ref{fig:fictitious} (a) and (d)).} \label{tab:title}
\centering
\begin{tabularx}{\columnwidth}{|Y|Y|Y|Y|Y|}
\hline
&    $\mathbf{t_1}$    &    $\mathbf{t_1^{\prime}}$   &    $\mathbf{t_2}$    &   $\mathbf{t_2^{\prime}}$   \\ \hline
Bulk      &   $-69.6$ &    $-69.5$ &    $68.8$  &    $68.9$  \\ \hline
Thin-Film &    $-67.6$ &    $-67.6$ &    $-60.4$ &    $-60.4$ \\ \hline
\end{tabularx}
\label{table:LUMO_hops}
\end{table}

\begin{figure*}[!htb]
    \centering
    \includegraphics[width = \textwidth]{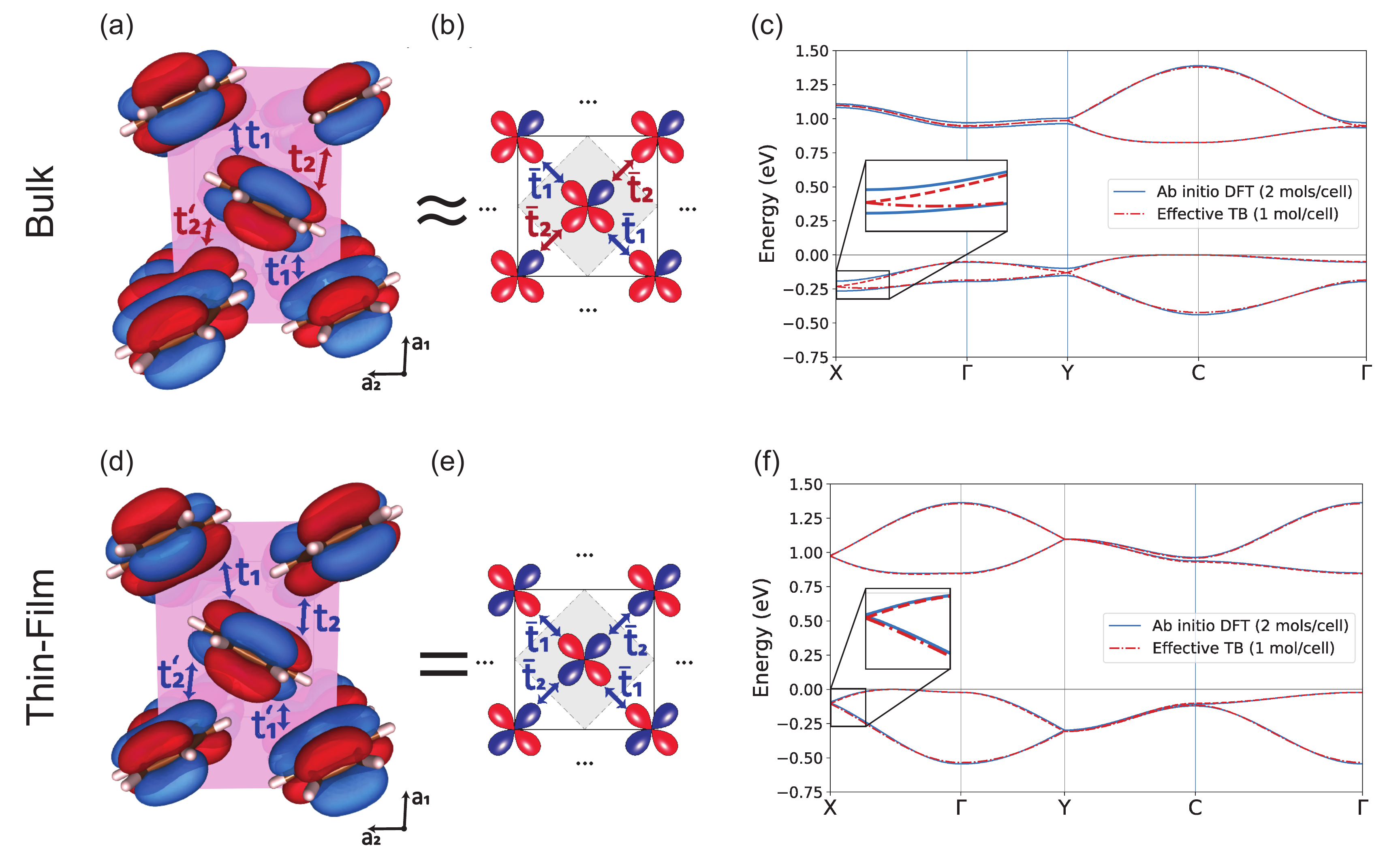}
    \caption{\label{fig:fictitious}(a), (d): Crystal structure of the two pentacene polymorphs along with the isosurface for the LUMO of the isolated molecules where red (blue) surfaces denote positive (negative) wavefunction sign. A pink plane perpendicular to the long axes of the molecules is shown to highlight the slipped-stacked nature of the bulk structure as compared to the thin-film. Arrows and $t$ denote the hopping integral between neighboring molecules in the basis of the frontier orbitals of isolated molecules. The nearly vanishing displacement of the molecules along their long axes in the thin-film stacking, which yields hopping integrals all with the same sign. (b), (e): LUMO orbital in the reduced unit cell (gray diamond) which yields the same symmetry of the hopping integrals as in (a), (d), consisting of averaged hopping parameters $\bar t_i = \frac{1}{2}(t_i + t_i')$. The orbitals shown here are fictitious, and only serve as a visual guide to represent the symmetry of the effective Hamiltonian. (c), (f): Effective tight-binding Hamiltonian band structure (red) with DFT-computed band structure (blue), showcasing the excellent accuracy of the effective Hamiltonian. The effective bands have been folded back into the BZ of the system with 2 molecules per unit cell. Note that the agreement between the DFT bands and effective picture is nearly perfect for thin-film pentacene, but deviates for bulk near the $X$ and $Y$ points on the zone edges. The original tight-binding band structure is plotted in the SI.}
\end{figure*}

The approximate equivalence of the hopping parameters $t_i$ and $t_i'$, which only differ by the sublattice index, motivates us to define a first-principles-backed effective Hamiltonian $\heff$ by averaging the tight-binding Hamiltonian matrix elements over the two sublattices,
\begin{align}
    \label{eqn:h_eff}
    h^{\text{eff}}_{ij}(\dtau) \equiv \frac{1}{2}\Big[h^{\vectau_1}_{ij}(\dtau) + h^{\vectau_2}_{ij}(\dtau)\Big].
\end{align}

The matrix elements in Eqn.~\ref{eqn:h_eff} no longer depend explicitly on any particular site index $\vectau$; hence, we interpret $\heff$ as a \emph{fictitious system with a single site per unit cell}. $\dtau$ becomes a lattice vector of this effective system, and $\vb{a}_{1,2}^\mathrm{eff} = \left(\vb{a}_1 \pm \vb{a}_2 \right)/2$ are the effective primitive lattice vectors (see Fig.~\ref{fig:fictitious}). We emphasize that this effective representation is valid for describing \emph{low-energy} excitations for bulk and thin-film pentacene, i.e., the two frontier electron and hole bands. Importantly, $\heff$ as above is only well-defined if $H$ has inversion symmetry.

Since $\heff$ was constructed only from inter-molecular hoppings, it contains no details of the molecular orbitals below this length scale; in particular, the effective orbitals $\chi_i^\mathrm{eff}$ of the fictitious system bear no resemblance with the original orbitals $\chi_i$. One could demand that effective HOMO and LUMO orbitals in $\heff$ should be derived within the same DFT Hamiltonian and yield the same hopping matrix elements as $H$. However, our results do not depend on the shape of the effective orbitals, only on the fact that they exist. Accordingly, we show a qualitative representation of such orbitals compatible with the symmetries of the hopping matrix elements in Fig.~\ref{fig:fictitious} (b) and (e). In the SI we detail our procedure to find high-symmetry orbitals that satisfy all hopping matrix elements, and show that it is always possible to construct them given $\heff$ and inversion in the original system.

In Fig.~\ref{fig:fictitious}~
(a) and (d), we show the unit cell of the original, physical system with two molecules per unit cell for the bulk and thin-film polymorphs, respectively, along with the LUMO orbital of an isolated molecule on each molecular site.

In panels (c) and (f), we show the quasipariticle band structure for both polymorphs computed with DFT and the effective tight-binding band structure obtained from $\heff$, folded back into the original Bruillion zone (BZ). The excellent agreement in $\heff$ reproducing the original DFT Hamiltonian stems from the HOMO and LUMO orbitals being highly localized, and displaying an internal symmetry within the unit cell. But despite the overall success of $\heff$, especially describing the thin-film polymorph, we note that it does not perfectly capture features such as the avoided crossings at the $X$ and $Y$ points for the bulk polymorph at the DFT level (insets in panels (c) and (f)). We now return to our discussion of singlet fission on the two polymorphs, and relate this small error to the singlet-fission rates.

Organic molecular crystals often support spin-triplet and -singlet excitons separated by a large exchange interaction. This condition is key to satisfy the singlet fission energy conservation condition that the energy of the lowest spin-singlet exciton, $E_S$, should be roughly twice that of the lowest spin-triplet exciton, $E_T$. There is a wide variety of different stable molecular packings that have been identified in experiment which satisfy this condition, including the two pentacene polymorphs studied here. However our new approach reveals that energy conservation alone is insufficient to guarantee efficient exciton-exciton coupling, and crystal packing can serve as a switch to turn on and off the singlet fission decay.

We compute from first-principles the exciton band structure in both polymorphs by solving the Bethe-Salpeter equation (BSE) for excitons with arbitrary center-of-mass (COM) momentum $\vb{Q}$~\cite{Deslippe2012,Qiu2015}. As shown in Fig.~\ref{fig:exciton_Vsb}, the exciton dispersion for both the singlet and triplet excitons are nearly identical in both polymorphs. With these quantities, we compute the singlet-fission transition rate using Fermi's golden rule\cite{Refaely-Abramson2017},
\begin{align}
    \tau^{-1} = \frac{2\pi}{\hbar}\frac{1}{\mathcal{V}_{xtal}}\sum_B \left|\braket{B|V_C|S} \right|^2\rho[\Delta E(S\rightarrow B)],
\end{align}
where $\mathcal{V}_{xtal}$ is the crystal volume, $V_C$ is the Coulomb operator, $\ket{S}$ is the initial, zero-momentum singlet exciton, $\ket{B}$ is the final, zero-momentum bi-triplet, and the density of states $\rho$ is taken to be a broadened delta function to account for the finite lifetime of the final states.

We focus on the singlet-fission coupling matrix element $\braket{B|V_C|S} \equiv \VSB$, describing the decay of the lowest-energy singlet with zero COM momentum to a bi-triplet via the Coulomb interaction $V_C$. The zero-momentum bi-triplet $\ket{B}$ is treated here as the product of two non-interacting but spin- and momentum-correlated triplet excitons. These two triplet states may have nonzero, but opposite, COM momenta, $\ket{B}\approx\ket{T,\vb{Q}}\ket{T',\vb{-Q}}$. The small coupling between the final triplets slightly changes the energy conservation rules, but does not change the selection rules we derive.

Figure~\ref{fig:exciton_Vsb} shows the singlet-fission coupling matrix elements $\VSB$ we compute as a function of the principal quantum numbers $T$ and $T'$ and COM momenta $\vb{Q}$ of the final bitriplet state $\ket{T,\vb{Q}}\ket{T',\vb{-Q}}$. It is evident that bulk pentacene supports nonzero $\VSB$ for large triplet COM momenta $\vb{Q}$, while, surprisingly, thin-film pentacene has much smaller $\VSB$ over the entire BZ. This is contrary to the na{\"i}ve expectation that the aforementioned similarities between the two polymorphs would imply similar singlet-fission coupling and rate.

The stark difference in the values of $\VSB$ between the bulk and thin-film polymorphs of pentacene is an indication of a hidden selection rule for exciton coupling which broadly affects exciton dynamics in molecular systems. These hidden selection rules can indeed be rationalized from the single-particle band structure and become apparent in the effective-Hamiltonian picture we derived, allowing us to explain the differences in singlet fission behavior at almost no extra computational expense.

\begin{figure}
    \centering
    \includegraphics[scale=.6]{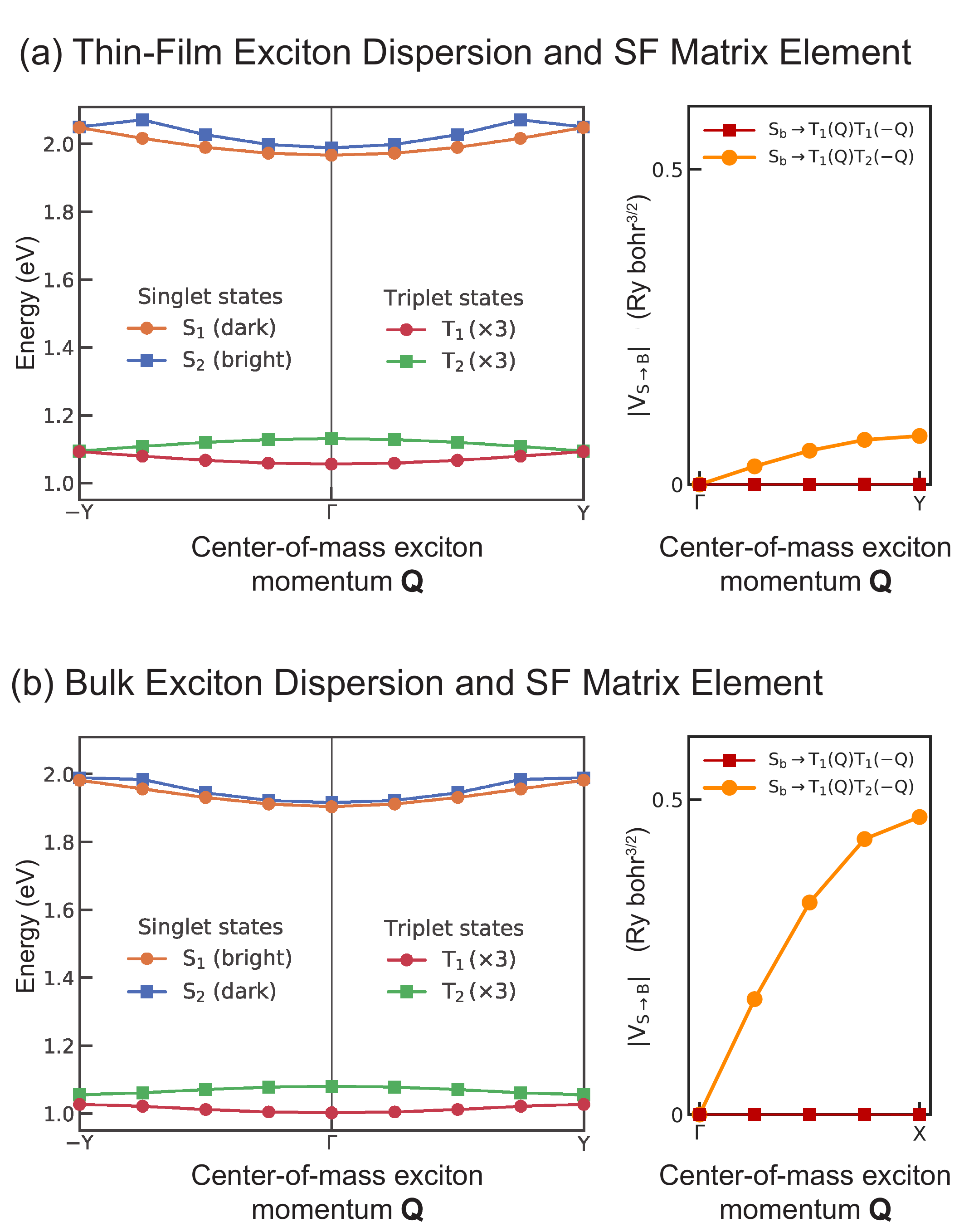}
    \caption{Left: Transverse exciton band structure computed with the \textit{ab initio} GW-BSE approach for (a) thin-film pentacene and (b) bulk pentacene. We show the four lowest-energy exciton states. Note the similarity in dispersion and energy between the polymorphs. Right: Singlet fission matrix element $|\VSB|$ for the transition from the bright singlet to either the same triplet band (red) or different triplet bands (orange) as a function of $\textbf{Q}$. The red curves are identically zero because of the inversion symmetry selection rule, while only the bulk has significant singlet fission for the transition onto triplet bands with different band index. The direction for the $\textbf{Q}$-cut in the BZ is chosen to maximize $\VSB$.}
    \label{fig:exciton_Vsb}
\end{figure}

We recall that a molecular crystal with a single molecule per unit cell displays a single low-energy triplet band. If such a system has inversion symmetry, Coulomb-mediated singlet fission is forbidden~\cite{Refaely-Abramson2017}, since
\begin{align}
\begin{split}
    \VSB &= \bra{T(\vb{Q})T(-\vb{Q})}\hat i ^{\dagger} V_{C} \hat i\ket{S(0)}= -\VSB,
    \label{eqn:sel_rule}
\end{split}
\end{align}
where we used the fact that the initial optically allowed singlet state is odd, and the final bitriplet is even under inversion symmetry ($\hat{i}$ is the inversion operator). In particular, the BSE Hamiltonian for the fictitious, high-symmetry Hamiltonian displays inversion symmetry and has one orbital per unit cell, hence \emph{$\heff$ does not support singlet fission}.

The selection rule in Eqn.~\ref{eqn:sel_rule} implies that it is not the variation of $\heff$ in different systems that determines the singlet fission rate. Instead, the governing quantity is the \emph{residual coupling} beyond the effective description of the effective Hamiltonian, defined by
\begin{align}
    \dV &= H - \heff.
\end{align}

In the bulk polymorph, $\dV$ is responsible for the splittings of $\sim50$~meV at the $X$ and $Y$ points in Fig.~\ref{fig:fictitious}~(c) not captured by $\heff$ -- not negligible compared to the valence-band bandwidth of $\sim500$~meV. Given the good agreement between $\heff$ and the DFT band structure, we note that $\dV$ is small in norm, but still responsible for the overall translation-symmetry breaking of the bulk structure beyond the description in $\heff$. On the other hand, in the thin-film structure, $\heff$ replicates the band structure almost perfectly and $\dV$ is almost negligible. We define a dimensionless figure of merit $F_{\Delta V}$ which approximates the coupling to allowed singlet fission matrix elements as a consequence of the translational-symmetry breaking from $\dV$ (see SI) as
\begin{align}
    F_{\Delta V} &\approx \ddfrac{2}{\delta D}\left(\overline{|\dV_e|} + \overline{|\dV_h|}\right), \label{fom}
\end{align}
where $\delta D$ is the difference between the second and third triplet bands (see SI), and $\overline{|\dV_{e/h}|}$ is the conduction/valence splitting averaged over the original BZ. In practice, $|\dV_{e/h}(\vb{k})|$ can be sampled at the avoided crossings of the electronic band structure computed within DFT or GW and is given by half the splitting at these points. Additionally, because $\delta D$ is related to the singlet-fission splitting and the exciton binding energy, it is relatively constant for different polymorphs and even different molecular crystals within the acene series, for instance~\cite{rangel2016structural}. Even without an accurate determination of $\delta D$ and, hence, the absolute value of $F_{\Delta V}$, we can straightforwardly compute the relative figure of merit between different polymorphs. For bulk and thin-film pentacene, we obtain $\left(F_{\Delta V}^{\mathrm{bulk}}\right)\big/\left(F_{\Delta V}^{\mathrm{thin}}\right) \approx 7.8$ within DFT, while within GW we obtain $\left(F_{\Delta V}^{\mathrm{bulk}}\right)\big/\left(F_{\Delta V}^{\mathrm{thin}}\right) \approx 15.8$. These ratios are consistent with the results in Fig.~\ref{fig:exciton_Vsb}, where bulk has approximately 12 times the singlet fission matrix element of thin-film. This excellent agreement shows that the analysis of the electronic and optical properties of organic molecular crystals in terms of this effective system with higher translational symmetry is indeed a powerful way not only to rationalize the features of the electronic dispersion of such materials, but also quantify the relative strength of processes involving multiexciton generation. The quantity $F_{\Delta V}$ is particularly useful for high-throughput calculations, as it allows the relative comparison of different singlet-fission materials at virtually no extra computational cost, and provides a more stringent criterion for determining SF beyond the singlet-bitriplet energy difference.

In conclusion, we develop a framework to analyze the electronic and optical spectrum of molecular crystals that naturally elucidates hidden symmetries associated with intracell and intermolecular interactions. Our approach finds the best single-site fictitious systems that yield the same low-energy excitations as the real, physical molecular crystal, and readily explains the difference in the quasiparticle band structure of two pentacene polymorphs -- bulk and thin-film. We also explain the origin of the much larger singlet-bitriplet Coulomb coupling in the bulk polymorph due to a small residual interaction that breaks the equivalence of the two molecular sublattices. Both the effective Hamiltonian and the symmetry breaking are obtained by simply analyzing the electronic band structure, and our conclusions are backed by first-principles many-body perturbation theory calculations. While we focused on two pentacene polymorphs as prototypical examples, our approach is not constrained by the number of neighbor molecules or the number of relevant orbitals per site. This strategy is also general to other excitonic processes, where a selection rule can be made stronger by considering it in the fictitious setting. We anticipate that this approach may also be used to identify which phonon modes are important for multiexciton processes by breaking the sublattice symmetry\cite{tempelaar2018vibronic, alvertis2019switching, alvertis2020impact, pandya2021exciton}. The newly proposed insight and selection rules provide a powerful computational tool and design paradigm for experiments and for high-throughput calculations to select materials with improved efficiency and functionality for excitonic processes.

\begin{acknowledgement}
ARA thanks Johnathan Georgaras and Supavit Pokawanvit for helpful discussions. SRA and FHJ thank Jeffrey B. Neaton and Steven G. Louie for stimulating discussions. This work was supported by the Center for Computational Study of Excited State Phenomena in Energy Materials (C2SEPEM), which is funded by the U.S. Department of Energy, Office of Science, Basic Energy Sciences, Materials Sciences and Engineering Division under Contract No. DE-AC02-05CH11231, as part of the Computational Materials Sciences Program. We acknowledge the use of computational resources at the National Energy Research Scientific Computing Center (NERSC), a DOE Office of Science User Facility supported by the Office of Science of the U.S. Department of Energy under Contract No. DE-AC02-05CH11231.
\end{acknowledgement}

\begin{suppinfo}

The following files are available free of charge.
\begin{itemize}
  \item Supplementary Information for Identifying Hidden Intracell Symmetries in Molecular Crystals and their Impact for Multiexciton Generation: Contains derivations for claims made in the main text, raw data used, calculation parameters, and extra figures.
\end{itemize}

\end{suppinfo}

\bibliography{bibliography}

\end{document}